\documentclass[english,aps, twocolumn, final]{revtex4}
\usepackage[T1]{fontenc}
\usepackage[latin9]{inputenc}
\usepackage{amstext}
\usepackage{graphicx}
\usepackage{esint}

\makeatletter
\@ifundefined{textcolor}{}
{%
 \definecolor{BLACK}{gray}{0}
 \definecolor{WHITE}{gray}{1}
 \definecolor{RED}{rgb}{1,0,0}
 \definecolor{GREEN}{rgb}{0,1,0}
 \definecolor{BLUE}{rgb}{0,0,1}
 \definecolor{CYAN}{cmyk}{1,0,0,0}
 \definecolor{MAGENTA}{cmyk}{0,1,0,0}
 \definecolor{YELLOW}{cmyk}{0,0,1,0}
}

\makeatother

\usepackage{babel}

\begin{document}





\title{Electromagnetic Polarizabilities of Mesons}

\author{A. Aleksejevs}
\address{Grenfell Campus of Memorial University, Corner Brook, Canada}

\author{S. Barkanova}
\address{Acadia University, Wolfville, Canada}



\begin{abstract}
The Chiral Perturbation Theory (CHPT) has been very successful in
describing low-energy hadronic properties in the non-perturbative
regime of Quantum Chromodynamics. The results of ChPT, many of which
are currently under active experimental investigation, provide stringent
predictions of many fundamental properties of hadrons, including quantities
such as electromagnetic polarizabilities. Yet, even for the simplest
hadronic system, a pion, we still have a broad spectrum of polarizability
measurements (MARK II, VENUS, ALEPH, TPC/2g, CELLO, Belle, Crystal
Ball). The meson polarizability can be accessed through Compton scattering,
so we can measure it through Primakoff reaction. This paper will provide
an analysis of the CHPT predictions of the SU(3) meson electromagnetic
polarizabilities and outline their relationship to the Primakoff cross
section at the kinematics relevant to the planned JLab experiments.
\end{abstract}




\maketitle

\section{Theory and Experiment to Date}

The study of hadronic structure with baryon-meson degrees of freedom
in the non-perturbative low energy regime of QCD has proved to be
rather successful with applications of Chiral Perturbation Theory
(CHPT) \cite{Ga84}. The role of CHPT in our understanding of hadron
structure has very broad spectrum of goals, ranging from the precise
calculations of hadron formfactors to the studies of polarizabilities.
Our attention goes to the polarizabilities, because their values clearly
reflect dynamical response of the mesons and baryons to the external
electromagnetic probe. To date our knowledge of the polarizabilities
is limited by the many experimental challenges, and only reliable
information is available for the proton's electric and magnetic polarizabilities:
$\alpha_{p}=(11.2\pm0.4)10^{-4}fm^{3}$ and $\beta_{p}=(2.5\pm0.4)10^{-4}fm^{3}$
\cite{PDG}. That is explained by the fact that polarizabilities are
accessible through the real or some form of virtual Compton scattering
and only experiments involving stable and charged target particle,
such as proton, have sufficient statistical significance. As a result,
polarizability of the neutron is less understood and plagued by the
large uncertainties: $\alpha_{n}=(11.6\pm1.5)10^{-4}fm^{3}$ and $\beta_{n}=(3.7\pm2.0)10^{-4}fm^{3}$
\cite{PDG}. As for the simplest $q\bar{q}$ state, such as meson
($P$), situation is even more complicated due to the meson short
lifetime $10^{-17}\sim10^{-8}$ sec. Up to now, only charged and neutral
pion polarizabilities have been measured, producing broad range of
values \cite{Ahrens,Antipov,Ba92}. In order to improve precision
of charged pion polarizability, at Jefferson Laboratory, GlueX collaboration
will measure polarizability through the Primakoff pion photo-production
reaction. Calculations of polarizabilities in CHPT also have been
completed for the pions up to order of $\mathcal{O}(p^{6})$ \cite{Ga06},
and for kaons up to order $\mathcal{O}(p^{4})$ in \cite{Guerrero}.
In this short paper we update calculations of the polarizabilities
for SU(3) octet of mesons using Computational Hadronic Model \cite{CHM}
with extension to lightest vector mesons. Moreover we produce analysis
of energy dependence of meson polarizabilities and study their impact
on the photon fusion cross sections $\sigma(\gamma\gamma\rightarrow P\bar{P})$
relevant to the Primakoff two-meson photo-production process.

\section{Meson Polarizability}

Meson electric and magnetic polarizabilities are determined from the
Compton structure functions $A(s,t)$ and $B(s,t)$ arising in the
Compton scattering: $\gamma+P\rightarrow\gamma+P$ 

\begin{eqnarray}
\alpha(s,t)=-\frac{1}{8\pi m}\bigg(A(s,t)+\frac{s-3m^{2}}{t}B(s,t)\bigg), & \text{and}\nonumber \\
\beta(s,t)=\frac{1}{8\pi m}\bigg(A(s,t)+\frac{s-m^{2}}{t}B(s,t)\bigg).\label{eq:1}
\end{eqnarray}

Polarizabilities represented in Eq.\ref{eq:1} are energy dependent
and hence we will call them dynamical. In the limit, then $s\rightarrow m^{2}$
and $t\rightarrow0$, we recover static values of the polarizabilities.
Compton structure functions from Eq.\ref{eq:1} are related to the
Compton tensor $M_{\mu\nu}=A(s,t)T_{\mu\nu}^{(1)}+B(s,t)T_{\mu\nu}^{(2)}$,
which enters Compton amplitude ($M=\epsilon'^{\mu}\epsilon^{\nu}M_{\mu\nu}$)
computable in CHM. Lorentz tensors $T_{\mu\nu}^{(1,2)}$ have the
following simple structure:

\begin{eqnarray}
&T_{\mu\nu}^{(1)}  =-\displaystyle{\frac{t}{2}g_{\mu\nu}-k_{3,\mu}k_{1,\nu}}\nonumber \\
\nonumber \\
T_{\mu\nu}^{(2)}& =\displaystyle{\frac{1}{2t}(s-m_{\pi}^{2})(u-m_{\pi}^{2})g_{\mu\nu}+k_{2,\mu}k_{2,\nu}+}\nonumber \\ 
\nonumber \\
&\displaystyle{\frac{s-m_{\pi}^{2}}{t}k_{3,\mu}k_{3,\nu}-\frac{u-m_{\pi}^{2}}{t}k_{2,\mu}k_{1,\nu}}.\label{eq:2} 
\end{eqnarray}

\begin{figure*}
\begin{centering}
\includegraphics[scale=0.65]{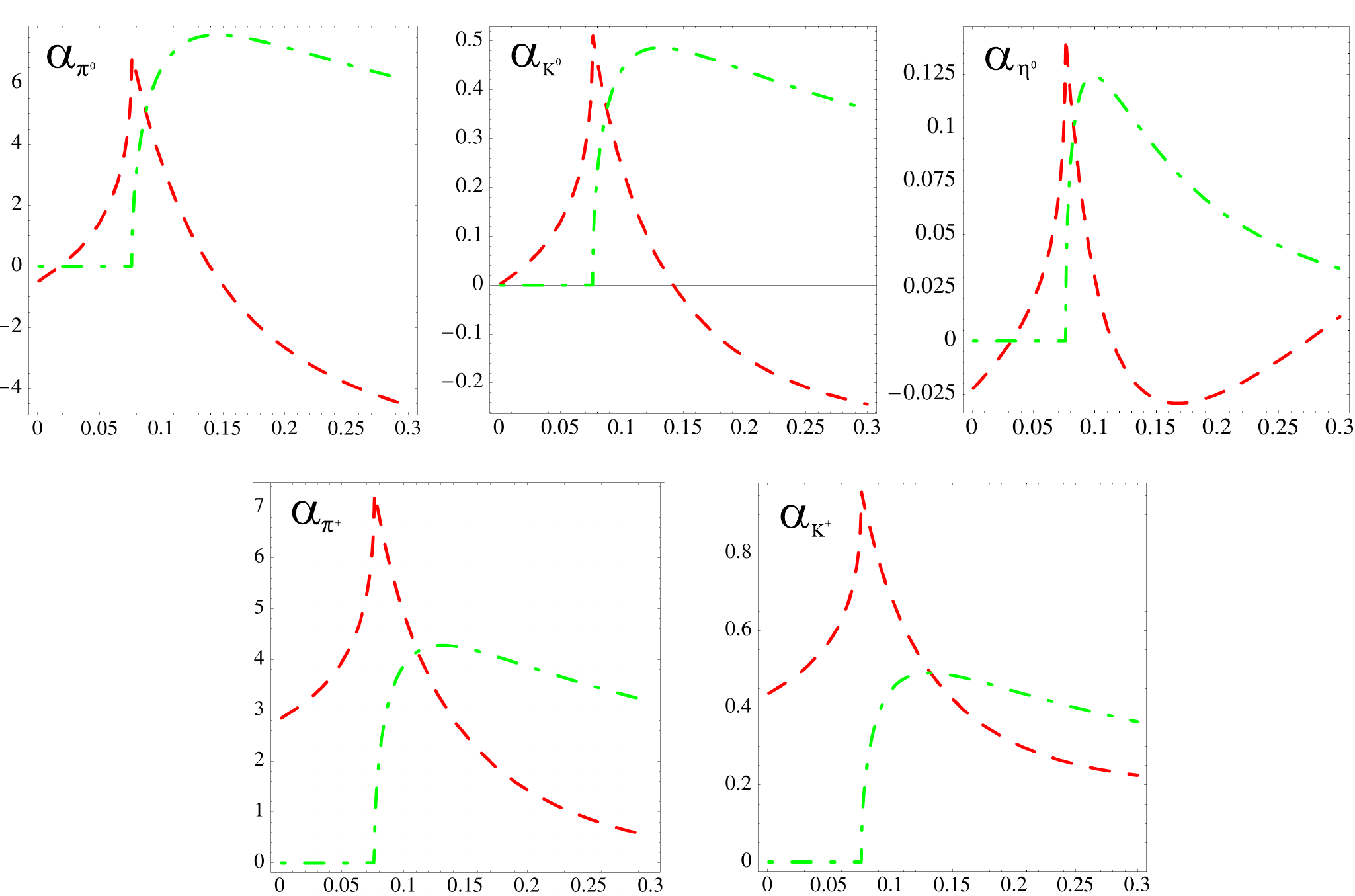}
\par\end{centering}

\caption{Energy dependencies of meson polarizabilities. The vertical axis on
all graphs represent electric polarizability in units of $10^{-4}\, fm^{3}$.
Horizontal axis shows center of mass energy squared $(s)$ in $GeV^{2}$.
Dashed (red) and dot-dashed (green) lines show $\text{Re}[\alpha_{P}(s,t\rightarrow-m_{P}^{2}]$
and $\text{Im}[\alpha_{P}(s,t\rightarrow-m_{P}^{2}]$ respectively.}

\label{fig1}
\end{figure*}

In Eq.\ref{eq:2}, $k_{\{1,3\}}$ and $k_{\{2,4\}}$ are momenta of
incoming/outgoing photon and meson respectively. The $s,t,$ and $u$
are the usual Mandelstam variables. Results for the static electric
and magnetic polarizabilities, produced with the help of CHM, can
be found in \cite{AB-Hadron-Structure-2013}. Since the experimental
values of the polarizabilities are extracted from the photon fusion
cross section $\sigma(\gamma\gamma\rightarrow\pi\pi)$ (see \cite{AB-Hadron-Structure-2013}),
it is more convenient to rotate Compton scattering from t- to s-plane
and use Eq.\ref{eq:1} with crossing symmetry $s\rightarrow t$ and
$t\rightarrow s$. In this case static values of the polarizabilities
are recovered from the following limits: $s\rightarrow0$ and $t\rightarrow-m_{P}^{2}$.
Using Eq.\ref{eq:1} and neglecting structure-dependent vector meson
pole contribution, in Fig.\ref{fig1}, we show results reproduced
by CHM for dynamical electric polarizabilities of mesons (here, we
have used $\alpha_{P}=-\beta_{P}$).

\section{Photon Fusion Cross Section in Meson Photoproduction}

In order to study impact of the polarizabilities on $\sigma(\gamma\gamma\rightarrow P\bar{P})$
cross section we recall the following expression \cite{Ba92,Do93,AB-Hadron-Structure-2013}:

\begin{eqnarray}
\sigma_{\gamma\gamma\rightarrow P\bar{P}}(|\cos\theta|<Z)=\frac{\kappa}{256\pi s^{2}}\int_{t_{a}}^{t_{b}}dt
\nonumber \\ \nonumber \\ \nonumber \\
\Bigg(\bigg|m_{P}^{2}B_{o}-8\pi sm_{P}\beta+\frac{4\pi}{m_{P}}(\alpha+\beta)st\bigg|^{2}+\label{eq:3} \nonumber \\
\nonumber \\ \nonumber \\
\bigg|B_{o}+\frac{4\pi s}{m_{P}}(\alpha+\beta)\bigg|^{2}\frac{(m_{P}^{4}-tu)^{2}}{s^{2}}\Bigg),
\end{eqnarray}

\begin{figure*}
\begin{centering}
\includegraphics[scale=0.65]{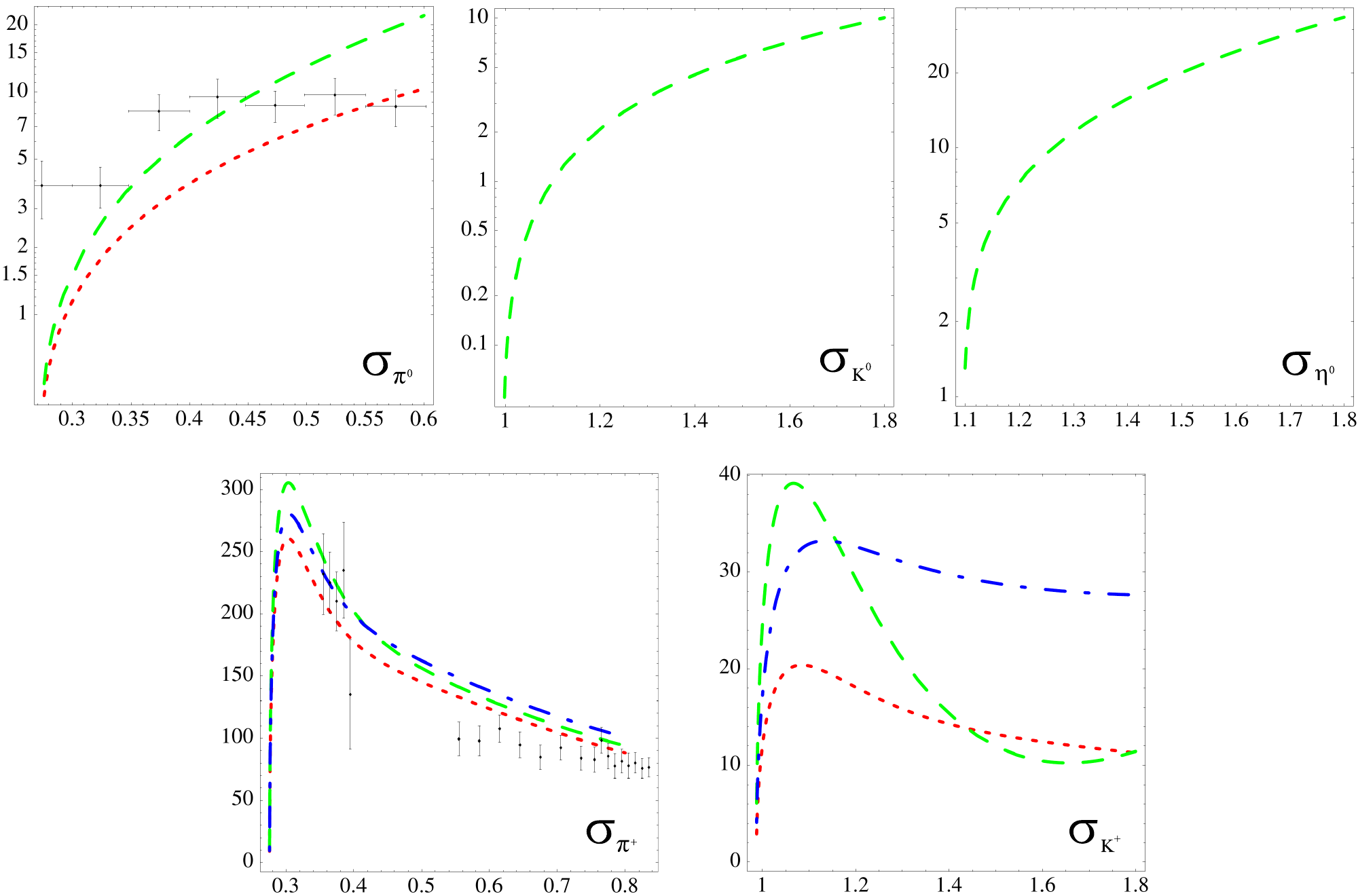}
\par\end{centering}

\caption{The $\sigma(\gamma\gamma\rightarrow P\bar{P})$ cross section dependency
on the invariant mass of produced mesons. The vertical axis shows
$\sigma_{\gamma\gamma\rightarrow P\bar{P}}(|\cos\theta|<0.6)\,(nb)$
and horizontal axis represents invariant mass of $P\bar{P}$ pair
in GeV. The dotted and red graph represent Born contribution. Dot-dashed
and blue graphs show cross section for the static polarizabilities
used in the Eq.\ref{eq:3}. The dashed and green graphs show results
for the cross section produced with dynamic polarizabilities from
Eq.\ref{eq:1}. For the charged pion polarizability data are taken
from MARK-II \cite{Bo92} and for neutral pion polarizability from
\cite{Crystall-Ball}.}

\label{fig2}
\end{figure*}

Here 
\begin{eqnarray*}
 & B_{o}=16\pi\alpha_{f}{\displaystyle \frac{s}{(t-m_{P}^{2})(u-m_{P}^{2})}}|q|,\\
\\
 & {\displaystyle t_{b,a}=m_{P}^{2}-\frac{1}{2}s\pm\frac{sZ}{2}\beta(s)},
\end{eqnarray*}
and $|q|$ stands for the charge of meson, $\beta(s)=\sqrt{\frac{s-4m_{P}^{2}}{s}}$
is the center of mass velocity of produced pair of mesons. Parameter
$\kappa=1$ or $2$ for the case of a neutral or charged meson, respectively.
Substituting polarizabilities from Eq.\ref{eq:1} to Eq.\ref{eq:3},
we show dependencies for the $\sigma(\gamma\gamma\rightarrow P\bar{P})$
cross section on invariant mass of the produced pair of mesons in
Fig.\ref{fig2}.

\section{Analysis and Conclusion}

First, it is quite noticeable that all meson polarizabilities on the
Fig.\ref{fig1} have rather strong energy dependence. Unlike baryon
dynamic polarizabilities \cite{Dyn-pol-Griesshammer-Hemmert,AB},
SU(3) meson electric and magnetic polarizabilities exhibit strong
excitation mechanism in the low energy domain and have similar slopes.
At the region of the pion production peak, electric polarizability
has the same resonance type shape for all SU(3) mesons. Same could
be said for the magnetic polarizability. At the order of $\mathcal{O}(p^{4})$
of CHPT, condition $\alpha_{P}(s,t)=-\beta_{P}(s,t)$ arises from
the fact that Compton structure function $B(s,t)$ is close to zero,
and only first significant non-zero contribution to $B(s,t)$ is derived
from two-loop $\mathcal{O}(p^{6})$ calculations. For the charged
pion, $\mathcal{O}(p^{6})$ calculations have been carried out in
\cite{Ga06} and it was shown that two-loop calculations have strong
impact on $\pi^{+}$ magnetic polarizability. It would be important
to see the $\mathcal{O}(p^{6})$ type of calculations for the polarizabilities
completed for the rest of SU(3) mesons and the analysis of their dynamical
behavior is carried out. As for the $\sigma_{\gamma\gamma\rightarrow P\bar{P}}(|\cos\theta|<0.6)\,(nb)$
cross sections, shown on the Fig.\ref{fig2}, it is clear that changes
in the charged pion polarizability from its static to the dynamic
value has only small impact on $\sigma_{\gamma\gamma\rightarrow\pi^{+}\pi^{-}}$
cross section (lower left plot on Fig.\ref{fig2}). This should certainly
provide some degree of model-independence when it comes to extraction
of the charged pion polarizabilities from Primakoff cross section.
For the rest of mesons, cross-sections are very sensitive to the variations
in the polarizability. For example, if we take cross section $\sigma_{\gamma\gamma\rightarrow K^{+}K^{-}}$
calculated with polarizability taken as static value (dot-dashed and
blue graph on the lower right part of Fig.\ref{fig2}) and compare
it to the cross section computed with dynamic polarizability (dashed
and green graph on the lower right part of Fig.\ref{fig2}), we can
see that difference is quite large for the entire region of $K^{+}K^{-}$
invariant masses. For the neutral mesons static polarizabilities are
close or equal to zero ($\alpha_{K^{0}}=0$) and hence difference
in cross sections is even more dramatic (we even do not show cross
sections for the neutral mesons computed with the static polarizabilities
because their values are close to zero). In general, we observe very
large sensitivity of the $\sigma_{\gamma\gamma\rightarrow P\bar{P}}$
cross sections (except for the charged pion) to the variations in
the meson polarizability. Obviously, meson polarizability calculated
in the different models will be strongly reflected in the different
shapes of the $\sigma_{\gamma\gamma\rightarrow P\bar{P}}$ cross section.
Essential interest goes to the charged kaon polarizability. Here,
we have second largest (after charged pion) photo-production cross
section and its shape is strongly correlated to the model in which
polarizability was calculated. It would be rather important to measure
energy dependence of $\gamma\gamma\rightarrow K^{+}K^{-}$ cross section
and extract charged kaon polarizability. This in turn should shed
some light on the effective models we are currently using in the low
energy QCD.

\nocite{*}

\begin{thebibliography}{10}
\bibitem{Ga84}J. Gasser and H. Leutwyler, Ann. Phys. 158, 142 (1984).

\bibitem{PDG}J. Beringer et al. (Particle Data Group), Phys. Rev.
D86, 010001 (2012).

\bibitem{Ahrens}J. Ahrens et al., Eur. Phys. J. A23, 113 (2005).

\bibitem{Antipov}Yu. M. Antipov et al., Phys. Lett. B121, 445 (1983).

\bibitem{Ba92}D. Babusci, et al. Phys. Lett. B 277, 158 (1992).

\bibitem{Ga06}J. Gasser, M.A. Ivanov, and M. E. Sainio, Nucl. Phys.
B745, 84 (2006).

\bibitem{Guerrero}F. Guerrero, J. Prades, Phys. Lett. B 405 (1997)
341.

\bibitem{CHM}A. Aleksejevs, M. Butler, J.Phys.G37,035002 (2010).

\bibitem{AB-Hadron-Structure-2013}A. Aleksejevs and S. Barkanova,
Nuclear Physics B (Proc. Suppl.) 245 (2013) 17\textendash{}24 

\bibitem{Do93}J. F. Donoghue and B. R. Holstein, Phys. Rev. D 48,
137 (1993).

\bibitem{Bo92}J. Boyer et al. (MARK-II collaboration), Phys. rev.
D 42, 1350 (1990).

\bibitem{Crystall-Ball}A.E. Kaloshin, V.V. Serebryakov, Phys. Lett.
B 278, (1992) 198-201.

\bibitem{Dyn-pol-Griesshammer-Hemmert}H. W. Griesshammer, T. R. Hemmert,
Phys. Rev. C 65, (2002) 045207.

\bibitem{AB}A. Aleksejevs and S. Barkanova, J.Phys. G38 (2011) 035004.

\end{thebibliography}


\end{document}